\documentclass[
pre,
reprint,
superscriptaddress,
showpacs,
nofootinbib,
amsmath,amssymb,
aps,
floatfix,
]{revtex4-1}

\usepackage{amsthm}
\usepackage{graphicx,color}
\usepackage{dcolumn}
\usepackage{bm}

\newcommand{\dd}{\mathrm{d}}

\newcommand{\GeV}{\mathrm{GeV}}
\newcommand{\MeV}{\mathrm{MeV}}
\newcommand{\fm}{\mathrm{fm}}

\newcommand{\ii}{\ensuremath{\mathrm{i}}}
\newcommand{\R}{\ensuremath{\mathbb{R}}}

\newcommand{\fB}{\ensuremath{f_{\text{B}}}}
\DeclareMathOperator\sign{sign}

\begin{document}

\title{Kinetic approach to a relativistic Bose-Einstein condensate}

\author{Alex Meistrenko}
\email{meistrenko@th.physik.uni-frankfurt.de}
\affiliation{Institut f{\"u}r theoretische Physik, Goethe-Universit{\"a}t Frankfurt am Main,
Max-von-Laue-Stra{\ss}e 1, 60438 Frankfurt, Germany}

\author{Hendrik van Hees}
\email{hees@fias.uni-frankfurt.de}
\affiliation{Institut f{\"u}r theoretische Physik, Goethe-Universit{\"a}t Frankfurt am Main,
Max-von-Laue-Stra{\ss}e 1, 60438 Frankfurt, Germany}
\affiliation{Frankfurt Institute for Advanced Studies,
  Ruth-Moufang-Stra{\ss}e 1, 60438 Frankfurt, Germany}

\author{Kai Zhou}
\email{zhou@th.physik.uni-frankfurt.de}
\affiliation{Institut f{\"u}r theoretische Physik, Goethe-Universit{\"a}t Frankfurt am Main,
Max-von-Laue-Stra{\ss}e 1, 60438 Frankfurt, Germany}
\affiliation{Frankfurt Institute for Advanced Studies,
  Ruth-Moufang-Stra{\ss}e 1, 60438 Frankfurt, Germany}

\author{Carsten Greiner}
\email{Carsten.Greiner@th.physik.uni-frankfurt.de}
\affiliation{Institut f{\"u}r theoretische Physik, Goethe-Universit{\"a}t Frankfurt am Main,
Max-von-Laue-Stra{\ss}e 1, 60438 Frankfurt, Germany}

 \date{\today}

\begin{abstract}
  We apply a Boltzmann approach to the kinetic regime of a relativistic
  Bose-Einstein condensate of scalar bosons by decomposing the
  one-particle distribution function in a condensate part and a non-zero
  momentum part of excited modes, leading to a coupled set of evolution
  equations which are then solved efficiently with an adaptive higher
  order Runge-Kutta scheme. We compare our results to the partonic
  cascade Monte-Carlo simulation BAMPS for a critical but far
  from equilibrium case of massless bosons. Motivated by the color glass
  condensate initial conditions in QCD with a strongly overpopulated
  initial glasma state, we also discuss the time evolution starting from
  an overpopulated initial distribution function of massive scalar
  bosons. In this system a self-similar evolution of the particle
  cascade with a non-relativistic turbulent scaling in the infrared
  sector is observed as well as a relativistic exponent for the direct
  energy cascade, confirming a weak wave turbulence in the ultraviolet
  region.
\end{abstract}


\maketitle

\section{Introduction}
\label{sec:1}

An ultracold atomic gas is a well-known and celebrated example of a
Bose-Einstein condensate (BEC) that arises close to the absolute zero
point of the temperature $T$ \cite{PhysRevLett.75.1687}:
\begin{equation} 
\begin{split}
f\left(E_i\right)=
\frac{1}{\exp\left(\frac{E_i-\mu}{T}\right)-1}\xrightarrow[T\to
0]{}0 \\
\text{for}\quad E_i>E_0\ge\mu\,.
\end{split}
\end{equation}
Thus in case of the Bose statistics the occupation number of all energy
states $E_i$ above the ground state $E_0$ becomes arbitrarily
small. Consequently, all particles of the system occupy the ground
state, leading to a macroscopic ground-state occupation number and
forming a coherent as well as a strongly correlated state. In
astrophysics it was proposed that dark matter, which is assumed to be a
cold bosonic and gravitationally bound system, could exist in the form
of a BEC \cite{1475-7516-2007-06-025}.  For compact stars the
condensation of mesons \cite{Menezes:2005ic} is discussed or even Cooper
pairing of baryons, which then can undergo a crossover transition to a
BEC state \cite{RevModPhys.80.1455}. Many more examples can be found in
condensed-matter theory, like condensation of magnons, which are
quasiparticles of spin waves. Magnons can condensate far from the
absolute zero point, to be more precise at room temperature
\cite{citeulike:876144}. Hence, there are two different possibilities
for the formation of a BEC state either by decreasing the temperature or
by increasing the particle density of bosons until it exceeds a critical
value. In this case the chemical potential becomes equal to the energy
of the ground state
\begin{equation} 
\begin{split}
f\left(E_0\right)=
\frac{1}{\exp\left(\frac{E_0-\mu}{T}\right)-1}\xrightarrow[\mu\to
E_0]{}\infty \\ 
\text{for}\quad E_0=m\ge\mu\,,
\end{split}
\end{equation}
leading once more to a macroscopic occupation number of $E_0$.  However,
in the thermodynamic limit of 3 spatial dimensions with
$\dd^3 p \sim p^2\dd p$ one does not obtain a macroscopic particle
number for $\mu \rightarrow m$ in the soft region by integrating the
momentum distribution function $f = \dd N/\dd^3 x \dd^3 p$ as seen from
the expansion $E=\sqrt{m^2+p^2} \simeq m+p^2/2m$. Thus one has to treat
the BEC in the kinetic regime as an extra contribution
$\sim \delta^{(3)}(\vec{p})$ to the one-particle distribution function
as will be discussed in Sec.\ \ref{sec:2}.

Since our motivation originates from high-energy physics, where we study
the chiral phase transition in the linear $\sigma$ model
\cite{vanHees:2013qla,Meistrenko:2013yya,Wesp:2014xpa,Greiner:2015tra},
we focus in Sec.\ \ref{sec:3} on initially overpopulated conditions
similar to the initial state of a heavy-ion collision. Due to
ultrarelativistic energies the two incoming heavy ions can be considered
as two slices of color glass condensate (CGC) which form a strongly
overpopulated initial glasma state during the collision process
\cite{McLerran:2001sr}. This state is highly dominated by gluons and
could lead on a short time scale to the formation of a condensate-like
state for gluons when in the kinetic regime binary scatterings dominate
over chemical processes \cite{Blaizot:2011xf, Blaizot:2013lga,
  Blaizot:2014jna, Scardina:2014gxa, Xu:2014ega, Huang:2013lia}. Such a
condensate state would be highly relativistic, since the relevant
saturation scale $Q_s$ is of the order of $1\;\GeV$, corresponding to
the square root of the initial number of partons per unit transverse
area. Thereby the occupation number is proportional to the inverse QCD
coupling constant $\alpha_s\left(Q_s\right)\ll 1$ up to the saturation
scale
\begin{equation}
f\left(t=0,p\right)\sim\frac{1}{\alpha_s}\quad\text{for}\quad p<Q_s\,,
\end{equation}
which suggests a perturbative treatment of the QCD in this limit.

In Sec.\ \ref{sec:2} we derive a set of evolution equations from the
general form of the Boltzmann equation and describe briefly our
numerical strategy for solving directly the partial integro-differential
equations. We then compare our results in Sec.\ \ref{sec:3} to the
partonic cascade simulation BAMPS (Boltzmann Approach to Multi-Parton
Scatterings) \cite{Xu:2004mz,Xu:2007aa}, which applies a stochastic test
particle ansatz to solve the collision integrals. Finally, we discuss
the dynamical evolution, starting from a strongly overpopulated initial
condition.

\section{Boltzmann approach with condensation}
\label{sec:2}

We start with the general expression for the time-dependent relativistic
Boltzmann-Uehling-Uhlenbeck equation
\begin{equation}
\begin{split}
  \partial_t f\left(t, \vec{p}_1\right) =
&\frac{1}{2E_1}\int\frac{\dd^3\vec
p_2}{\left(2\pi\right)^32E_2}\int\frac{\dd^3\vec{p}_3}{\left(2\pi\right)^32E_3}
\int\frac{\dd^3\vec{p}_4}{\left(2\pi\right)^32E_4} \\
&\times\left(2\pi\right)^4\delta^{(4)}\left(P_1+P_2-P_3-P_4\right)\frac{\left |\mathcal{M}_{12\rightarrow 34} \right|^2}{\nu}
 \\ &\times
[\left(1+f_1\right)\left(1+f_2\right)f_3f_4 \\ & \quad
-f_1f_2\left(1+f_3\right)\left(1+f_4\right) ]\,,
\label{eq:200}
\end{split}
\end{equation}
where $f_i:=f\left (t,\vec{p}_i\right )$ with $i\in\{1,2,3,4\}$ denotes
the one-particle distribution function, $\nu$ is the degeneracy factor,
and $P_i$ stands for the particle's on-shell four-momentum,
$P_i^0=E_{\vec{p}_i}=\sqrt{m_i^2+\vec{p}_i^2}$. Following similar lines
as in reference \cite{PhysRevLett.74.3093} we decompose the distribution
function in two parts,
\begin{equation}
\begin{split}
f\left(t,\vec{p}\right)&=f_{\text{part}}\left(t,\vec{p}\right)+f_{\text{cond}}\left(t,\vec{p}=0\right)\\
&=f_{\text{part}}\left(t,\vec{p}\right)+n_{\text{cond}}\left(t\right)
\left (2 \pi\right)^3 \delta^{(3)}(\vec{p})\,.
\label{eq:201}
\end{split}
\end{equation}
The first expression $f_{\text{part}}\left(t,\vec{p}\right)$ denotes the
distribution function of non-zero momentum modes (particles), whereas
the second one $f_{\text{cond}}\left(t,\vec{p}=0\right)$ contains the
condensate contribution at zero momentum. The latter part is
proportional to the condensate density $n_{\text{cond}}\left(t\right)$
and a $\delta$-function in momentum space.  By inserting the
decomposition \eqref{eq:201} in \eqref{eq:200} and comparing the
coefficients, one derives the evolution equation for the condensate,
\begin{equation}
\partial_t f_{\text{cond}}\left(t, \vec{p}_1=0\right)=\mathcal
C_{1\text{c}+1\text{p}\leftrightarrow 2\text{p}} \, ,
\label{eq:202a}
\end{equation}
with the condensate-particle collision term
\begin{equation}
\begin{split} \mathcal C_{1\text{c}+1\text{p}\leftrightarrow
2\text{p}}=&\frac{1}{2E_1}\int\frac{\dd^3\vec{p}_2}{\left(2\pi\right)^32E_2}
\int\frac{\dd^3\vec{p}_3}{\left(2\pi\right)^32E_3}\int\frac{\dd^3\vec{p}_4}{\left(2\pi\right)^32E_4}
\\ &\times\left(2\pi\right)^4\delta^{(4)}\left (P_1+P_2-P_3-P_4\right
)\frac{\left |\mathcal M_{12\rightarrow 34} \right|^2}{\nu} \\
&\times\left[f_{\text{c}}f_3f_4-f_{\text{c}}f_2\left(1+f_3+f_4\right)\right]\,.
\label{eq:202b}
\end{split}
\end{equation}
The evolution equation for the non-zero modes of the distribution
function consists of two collision integrals,
\begin{equation}
\partial_t f_{\text{part}}\left(t, \vec{p}_1\neq 0\right) = \mathcal
C_{2\text{p}\leftrightarrow 2\text{p}}+\mathcal
C_{1\text{p}+1\text{c}\leftrightarrow 2\text{p}}\,,
\label{eq:203a}
\end{equation}
which include the pure particle as well as the particle-condensate
interactions,
\begin{equation}
\begin{split} \mathcal C_{1\text{p}+1\text{c}\leftrightarrow
2\text{p}}=&\frac{1}{2E_1}\int\frac{\dd^3\vec{p}_2}{\left(2\pi\right)^32E_2}\int\frac{\dd^3\vec{p}_3}{\left(2\pi\right)^32E_3}\int\frac{\dd^3\vec{p}_4}{\left(2\pi\right)^32E_4}\\
&\times\left(2\pi\right)^4\delta^{(4)}\left (P_1+P_2-P_3-P_4\right
)\frac{\left |\mathcal M_{12\rightarrow 34} \right |^2}{\nu}\\
&\times\Big\{\left[f_{\text{c}}f_3f_4-f_{\text{c}}f_1\left(1+f_3+f_4\right)\right]\\
&\qquad+\left[\left(1+f_1+f_2\right)f_{\text{c}}f_4-f_{\text{c}}f_1f_2\right]\\
&\qquad+\left[\left(1+f_1+f_2\right)f_{\text{c}}f_3-f_{\text{c}}f_1f_2\right]\Big\}\,.
\label{eq:203b}
\end{split}
\end{equation}
Due to energy-momentum conservation the condensate part appears only
once per gain and loss term of expressions \eqref{eq:202b} and
\eqref{eq:203b}. The collision integral
$\mathcal C_{2\text{p}\leftrightarrow 2\text{p}}$ has the same structure
as the right-hand side of \eqref{eq:200}, including only non-zero modes.

The evolution of the condensate is fully described by the condensate
density since the $\delta$-function can be integrated out
\begin{equation}
\begin{split}
\partial_t
n_{\text{cond}}\left(t\right)&=\int\frac{\dd^3\vec{p}_1}{\left(2\pi\right)^3}\partial_t
f_{\text{cond}}\left(t, \vec{p}_1=0\right)\\
&=\int\frac{\dd^3\vec{p}_1}{\left(2\pi\right)^3}\mathcal
C_{1\text{c}+1\text{p}\leftrightarrow
2\text{p}} \\
&=-\int\frac{\dd^3\vec{p}_1}{\left(2\pi\right)^3}\mathcal
C_{1\text{p}+1\text{c}\leftrightarrow 2\text{p}}\,.
\label{eq:204}
\end{split}
\end{equation}
Because of the last relation the total particle density is conserved,
and particles with zero momentum can be interpreted as constituent parts
of the condensate.

For a first study of BEC dynamics we focus on an isotropic system of
scalar bosons with $\phi^4$-interaction\footnote{We choose
  $\mathcal{L}_{\text{int}}=\frac{\lambda}{4}\phi^4$ as the quartic
  interaction term in the Lagrangian.}
\begin{equation} \frac{1}{\nu}|\mathcal M_{12\rightarrow
34}|^2=18\lambda^2\,.
\label{eq:205}
\end{equation}
In this case the evolution equations \eqref{eq:202a} as well as
\eqref{eq:203a} simplify and have the form given by \eqref{eq:a11},
\eqref{eq:a12}.

The growth of the condensate happens exponentially fast as can be seen
from the form of \eqref{eq:a12}. This requires an efficient numerical
solution of the integro-differential equations. Therefore we use an
adaptive numerical scheme and apply the following steps.
\begin{enumerate}
\item We introduce a non-uniform discretization for the external
  $p_1$-momentum with decreasing resolution from the soft to the hard
  region of momenta, such that the distribution function is now given on
  a large grid $G:=\{p^0,p^1,...,p^N\}$ with $N>100$. Thereby for the
  soft discrete momenta the relation $0 <p^1<...< p^k< m$ with $k\gg 1$
  holds. That fulfills a mandatory condition in the infrared,
  reproducing the right scaling close to equilibrium as shown later in
  Eq.\ \eqref{eq:305b}.
\item We note that according to the form of
  \eqref{eq:a12} the onset of the condensation process can only happen for
  a finite initial value of $n_\text{c}\left(t=0\right)$. It was shown 
  in \cite{Epelbaum:2014mfa} that the time evolution is independent of the 
  initial value as long as $n_\text{c}\left(t=0\right)$ is negligibly small 
  compared to the total particle density $n_\text{tot}$ of the system. Since 
  the condensate starts to grow rapidly when 
  the effective chemical potential\footnote{Obtained from a fit to the Bose
  distribution function \eqref{eq:304}.} of the system
  becomes equal to the boson mass, it is also possible to insert an initial
  seed for $n_\text{c}$ at this time point. Both methods result in almost equal
  time scales for the condensation process.
  
  We solve the isotropic two-dimensional collision integrals in
  \eqref{eq:a11} and \eqref{eq:a12} for every external mode by applying
  quadrature methods.
  The single integrals for the external momentum modes are independent of each other
  and we can solve them in parallel. Due to the fact that $p_4$ is
  integrated out with the energy $\delta$-function, in general it does
  not lie exactly on the grid.  For this momentum the distribution
  function is interpolated by using cubic splines on the discrete values
  of the grid $G$.

  When needed also Monte Carlo importance sampling
  \cite{PeterLepage1978192} can be used to calculate the collision
  integrals. In this case the distribution function is interpolated for
  all internal momenta $p_2, p_3$ and $p_4$.
\item Finally, the differential equations themselves are solved with the
  Cash-Karp RK45-scheme \cite{Cash:1990:VOR:79505.79507} that guarantees
  an adaptive slowing down of the simulation during the condensation
  process and a fast convergence due to the $5^{\text{th}}$-order
  accuracy with respect to the time discretization $\Delta t$ of the
  scheme.
\end{enumerate}
A different implementation of a discrete Boltzmann equation for scalar
particles in $(2+1)$ dimensions and overpopulated systems can be found
in \cite{Epelbaum:2015vxa}. Here, the authors applied their code to a
longitudinally expanding system with CGC-like initial conditions, which
mimics the kinematics of a heavy ion collision at very high energies.

\section{Thermalization in overpopulated systems}
\label{sec:3}

As discussed in Sec.\ \ref{sec:1} Bose-Einstein condensation can arise
in ultracold or overpopulated systems. Motivated by the color-glass
initial conditions in ultrarelativistic heavy-ion collisions, we focus
on an isotropic step function as initial momentum distribution and
consider the case of overpopulation for scalar bosons
\begin{equation} f\left (t=0,p\right )=f_0\,\theta\left
(1-\frac{p}{Q_s}\right )\,.
\label{eq:301}
\end{equation}
Here $Q_s= 1 \; \GeV$ denotes the saturation scale, and
$f_0\sim\mathcal{O}\left (1\right )$ is a constant\footnote{Typically, in the gluon
saturation regime one obtains $f_0\approx 2-4$.}. The total particle as well as energy densities are then given by
\begin{equation}
\begin{split} 
n_{\text{tot}}&=\frac{f_0Q_s^3}{6\pi^2}\,,\\
\epsilon_{\text{tot}}&=\frac{f_0}{16\pi^2}\Bigg [Q_sE_{Q_s}\left
(m^2+2Q_s^2\right ) \\
&\qquad +m^4\log\frac{m}{Q_s+E_{Q_s}}\Bigg ]
\label{eq:302}
\end{split}
\end{equation}
with $m$ being the mass of the bosons and
$E_{Q_s}=\sqrt{m^2+Q_s^2}$. Starting with a nonequilibrium initial
condition the system has to end up in an equilibrium distribution
function, to be more precise in a Bose distribution. However, in an
overpopulated and particle conserving system it will be in general not
possible to fit the total particle and energy density at the same
time. Consequently, such a system requires that a macroscopic number of
particles populates the ground state, i.e., the formation of a
Bose-Einstein condensate becomes necessary to reach an equilibrium
state. The time evolution of the condensate density is directly related
to the evolution of the one-particle distribution function by the
following relations:
\begin{equation}
\begin{split} n_{\text{cond}}\left (t\right
)&=n_{\text{tot}}-n_{\text{part}}\left (t\right )\,,\\
n_{\text{part}}\left (t\right )&=\int\frac{\dd^3\vec
p}{\left(2\pi\right)^3}f_{\text{part}}\left(t,\vec{p}\right)\,,\\\\
\epsilon_{\text{cond}}\left (t\right
)&=\epsilon_{\text{tot}}-\epsilon_{\text{part}}\left (t\right
)\,,\\
\epsilon_{\text{part}}\left(t\right)&=\int\frac{\dd^3\vec
p}{\left(2\pi\right)^3}Ef_{\text{part}}\left(t,\vec{p}\right)\,.
\label{eq:303}
\end{split}
\end{equation}
At thermal equilibrium this set of equations can be solved
self-consistently, leading to the Bose distribution as the equilibrium
distribution function of non-zero modes:
\begin{equation}
\begin{split}
\fB\left(E\right)=\frac{1}{\exp \left (\frac{E-\mu}{T} \right)-1}\\
\text{with} \quad E=\sqrt{p^2+m^2}\,.
\label{eq:304}
\end{split}
\end{equation}
In case of BEC formation the chemical potential in \eqref{eq:304}
becomes $\mu=m$. So the only unknowns are the critical temperature
$T_{\text{crit}}$ and the condensate density $n_{\text{cond}}$, since
the energy density is given by the relation
$\epsilon_{\text{cond}}=m\,n_{\text{cond}}$.
\begin{figure}
 \includegraphics[width=\columnwidth]{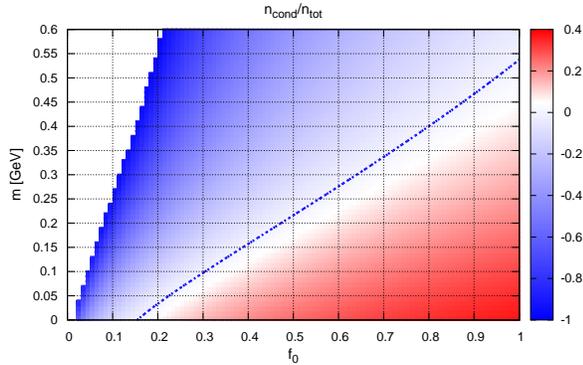} \centering
 \caption{Numerical solution of \eqref{eq:303} for the condensate
   fraction $n_{\text{cond}}/n_{\text{tot}}$ under the assumption of
   $\mu = m$ as a function of the mass $m$ and the step height
   $f_0$. Below the blue line a Bose-Einstein condensate forms. The
   region above the blue line is unphysical since the condition
   $\mu = m$ does not apply anymore.}
 \label{fig:301}
\end{figure}
Fig.\ \ref{fig:301} shows the numerical solution of \eqref{eq:303} for
the condensate fraction $n_{\text{cond}}/n_{\text{tot}}$ in equilibrium
as a function of the mass $m$ and the step height $f_0$. A condensate
formation occurs below the blue line, which marks the critical case of
$n_{\text{part}}=n_{\text{tot}}$. The unphysical region of negative
values for $n_{\text{cond}}$ above the blue line results from the
assumption $\mu = m$, which was applied to solve \eqref{eq:303}.

\subsection{Dynamics of the critical case}
\label{sec:31}

Before considering strongly overpopulated systems for massive scalar
bosons, we compare the dynamical evolution of the critical case with the
well-established partonic transport code BAMPS, which has been developed
for gluon and quark transport in heavy-ion collisions. It employs a
test-particle ansatz to approximate the phase-space distribution and
then solves the collision integrals via a stochastic interpretation
\cite{Xu:2004mz,Xu:2007aa}. We note that BAMPS treats gluons as massless
bosons, and therefore close to equilibrium the momentum distribution
function shows the following scaling in the infrared regime
\eqref{eq:304}:
\begin{figure}
 \includegraphics[width=\columnwidth]{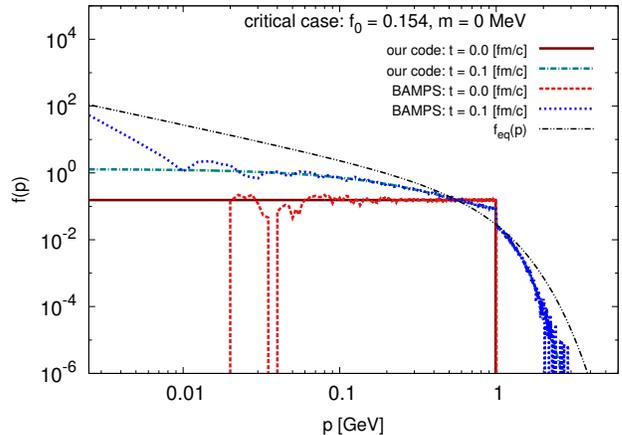}
\centering
\caption{Momentum distribution function at $t=0.1\;\text{fm}/c$,
  starting from a step function with $f_0=0.154$, $m=0\;\MeV$.  For
  comparison also BAMPS results (from a single run) of massless bosons
  are plotted.}
 \label{fig:302}
\end{figure}
\begin{figure}
 \includegraphics[width=\columnwidth]{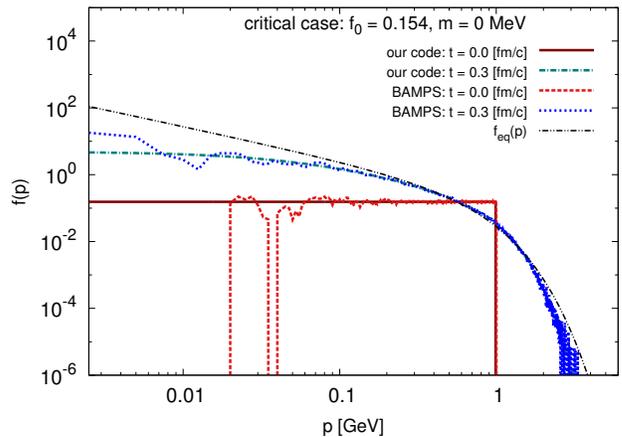}
\centering
\caption{The same as Fig. \ref{fig:302} but for $t=0.3\;\text{fm}/c$}
 \label{fig:303}
\end{figure}
\begin{figure}
 \includegraphics[width=\columnwidth]{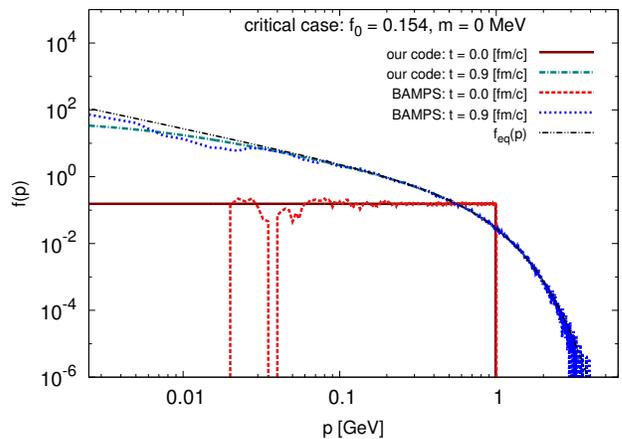}
\centering
\caption{The same as Fig. \ref{fig:302} but for $t=0.9\;\text{fm}/c$
  with the system being close to the equilibrium state.}
 \label{fig:304}
\end{figure}
\begin{equation}
f\left(p\right)\sim\frac{T}{p}\quad\mbox{for}\quad p\ll T\,,
\label{eq:305a}
\end{equation}
whereas in the massive case one obtains
\begin{equation}
f\left(p\right)\sim\frac{2mT}{p^2}\quad\mbox{for}\quad p\ll m\ll T\,.
\label{eq:305b}
\end{equation}
Thus in the very soft region the momentum distribution function of
massless bosons behave differently in comparison to massive bosons.

For the numerical comparison we initialize a single BAMPS run with a
cross section equivalent to the transition amplitude of the
$\phi^4$-theory \eqref{eq:205} with a relatively large coupling constant
$\lambda\approx 46.4$ which reproduces the typical thermalization scale
of QCD. According to Eqs.\ \eqref{eq:a11}, \eqref{eq:a12} a massless
approach like BAMPS is applicable for a vanishing value of the
condensate\footnote{Consequently, a finite value of the condensate
requires massive bosons for the simple transition amplitude of the
$\phi^4$-theory \eqref{eq:205}.}. Also in our simulation we can handle
the massless case, noting that now the numerical condition in the soft
region of discrete momenta has to fulfill $0 <p^1<...<p^k< T$ with
$k\gg 1$ for being able to reproduce the correct infrared behavior as
seen from the form of \eqref{eq:305a}.

Figs. \ref{fig:302}-\ref{fig:304} show the evolution of the momentum
distribution functions for both simulations at three intermediate times
between the initial and final states. In the critical case of massless
bosons the initial step height is $f_0 = 0.154$ as can be seen from
Fig. \ref{fig:301}.  The dynamics of our simulation is in good agreement
with the BAMPS calculation.  Only in the soft region the evolution
differs due to a lack of statistics in a single run of BAMPS.

\subsection{Dynamics of the strongly overpopulated state}
\label{sec:32}

\begin{figure}
 \includegraphics[width=\columnwidth]{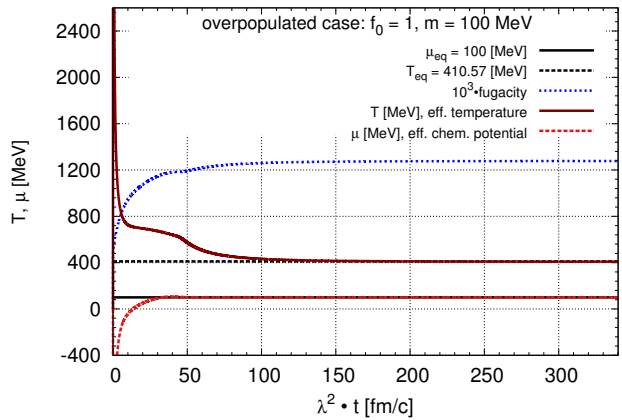}
\centering
 \caption{Evolution of the effective temperature $T$, chemical potential
$\mu$ and fugacity as a function of $\lambda^2t$ for $f_0=1$,
$m=100\; \MeV$.}
 \label{fig:305}
\end{figure}

\begin{figure}
 \includegraphics[width=\columnwidth]{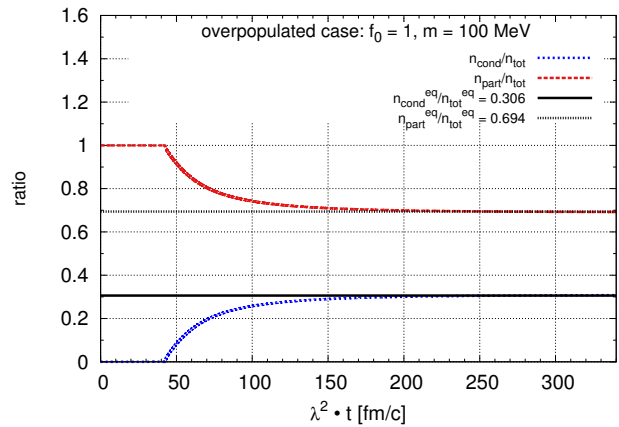}
\centering
\caption{Evolution of the condensate density $n_{\text{cond}}$ with
  respect to the total particle density $n_{\text{tot}}$ as a function of
  $\lambda^2t$ for $f_0=1$, $m=100 \; \MeV$.}
 \label{fig:306}
\end{figure}

\begin{figure}
 \includegraphics[width=\columnwidth]{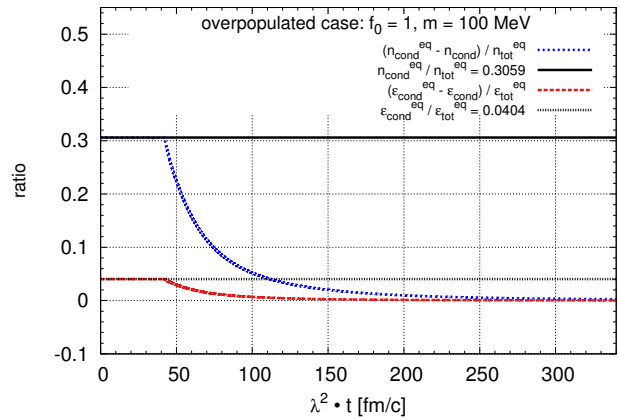}
\centering
\caption{Relative deviation from equilibrium values of the condensate
  density $n_{\text{cond}}$ and energy density $\epsilon_{\text{cond}}$
  as a function of $\lambda^2t$ for $f_0=1$, $m=100\; \MeV$.}
 \label{fig:307}
\end{figure}

We initialize a strongly overpopulated system according to
\eqref{eq:301} with $f_0=1$, $Q_s=1\;\GeV$ and a boson mass of
$m=100\;\MeV$.  Since the thermalization scale is simply proportional to
$1/\lambda^2$ all results are plotted independently of the coupling
constant with $\lambda^2t$ on the abscissa. Fig.\ \ref{fig:305} shows the
time evolution for the effective temperature and chemical potential as
well as the fugacity defined as $z:=\exp\left(\mu/T\right)$.
The onset of condensation happens at around
$\lambda^2t\approx 40\;\text{fm}/c$ when the chemical potential in the
soft region shortly overshoots the equilibrium value of
$\mu_{\text{eq}}=100\; \MeV$ and then converges rapidly to
$\mu_{\text{eq}}$. Also the effective temperature converges exactly to
the expected value of $T_c\approx 410.57\; \MeV$ leading to
$z\approx 1.28$. In addition Figs.\ \ref{fig:306} and \ref{fig:307} show
the evolution of different ratios for the quantities $n_{\text{cond}}$, 
$n_{\text{part}}$ as well as $\epsilon_{\text{cond}}$, $\epsilon_{\text{part}}$. 
These plots underline the conservation of the total densities $n_{\text{tot}}$, 
$\epsilon_{\text{tot}}$ to high accuracy.
\begin{figure}
 \includegraphics[width=\columnwidth]{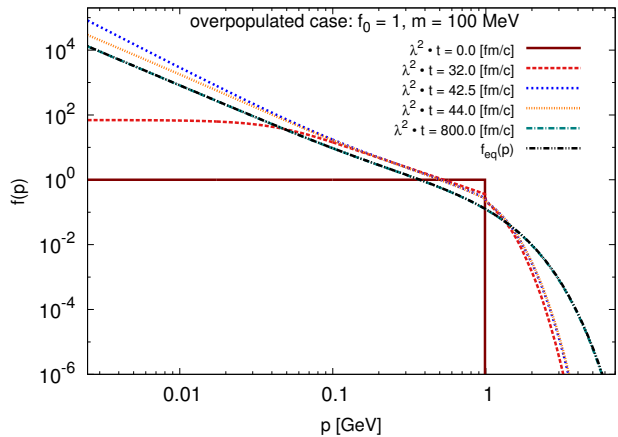}
\centering
\caption{Time evolution of the one-particle momentum distribution
  function, starting from an overpopulated initial state, with respect
  to the coupling-independent time scale $\lambda^2t$ for $f_0=1$ and
  $m=100\; \MeV$.}
 \label{fig:308}
\end{figure}

\begin{figure}
 \includegraphics[width=\columnwidth]{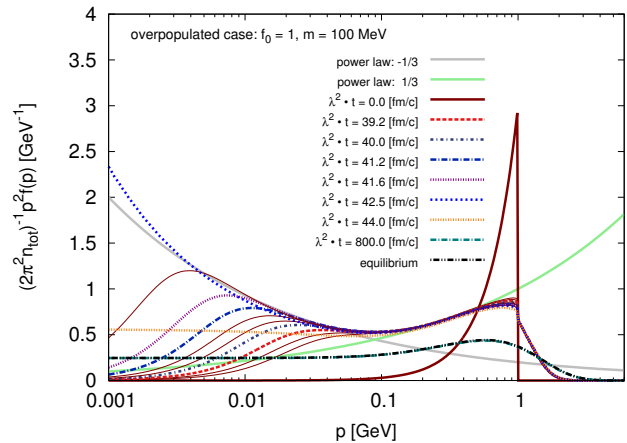}
\centering
\caption{Differential fraction of the isotropic particle density with
  respect to the momentum, showing different power-law scalings in the
  region of relativistic and non-relativistic momenta.}
 \label{fig:309b}
\end{figure}

\begin{figure}
 \includegraphics[width=\columnwidth]{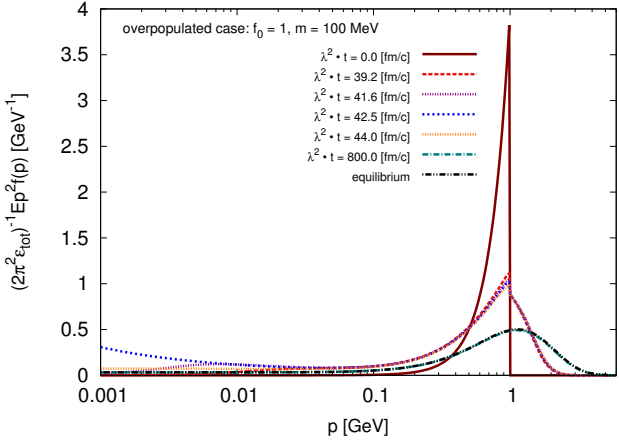}
\centering
\caption{Differential fraction of the isotropic energy density with respect to the momentum.}
 \label{fig:309c}
\end{figure}

Now, we consider the time evolution of the one-particle distribution
function. Starting from such a strongly overpopulated state the
distribution function grows by $5$ orders of magnitude for
$p=2$-$3\; \MeV$ and overshoots significantly the equilibrium
distribution function. The reason is a strong particle cascade to the
infrared, resulting in an inverse energy flow to the ultraviolet region
as will be discussed in the following. Due to the onset of the
condensation process around $\lambda^2t\approx 40\;\text{fm}/c$ the
particle number of the soft momenta starts to decrease and converges
rapidly to the equilibrium shape, whereas the hard modes develop slowly
in time, resulting in a large equilibration time compared to the time
scale of the condensation. As already seen in Fig.\ \ref{fig:305} for
the effective temperature and chemical potential, the evolution of the
system reaches the equilibrium state as a stable fixed point. In the
equilibrium state a finite value of the condensate coexists with a Bose
distribution function at non-zero momenta. Thereby the condensate value
matches exactly the expected value of $n_{\text{cond}}=0.673\;\fm^{-3}$
which once more underlines the accurate convergence of our kinetic
approach (compare also Fig.\ \ref{fig:309b} and \ref{fig:309c}).

To discuss the one-particle distribution function in detail we focus on
references \cite{PhysRevLett.74.3093,Semikoz:1995rd}, where the authors
studied the evolution of a similar system in a classical approach with
the dispersion relation $E=p^2/2m$ and analyzed the energy scaling in
detail. Thereby they observed a self-similar evolution in time before
the onset of the condensation process and obtained a particular
power law $f(E)\sim 1/E^{\alpha}$ with the numerical exponent
$\alpha_{\text{num}}\approx 1.24$ in the deep infrared region, which was close
to the value $\alpha=7/6$. It was pointed out by the authors that the
exponent $\alpha$ marks a non-thermal fixed point for the distribution
function and acts as a root for the collision integral of the classical
Boltzmann equation as can be shown by means of a Zakharov
transformation\footnote{Further references can be found in
  \cite{Semikoz:1995rd}.}. Consequently, there exists an energy scale,
where the occupation number changes only slowly in time. Due to particle
and energy conservation a self-similar evolution of this energy scale
can be observed, growing in time and propagating towards the infrared.
To be more precise they found a turbulent transport of scalar bosons to
the infrared, which is known as an inverse particle cascade.
\begin{figure}
 \includegraphics[width=\columnwidth]{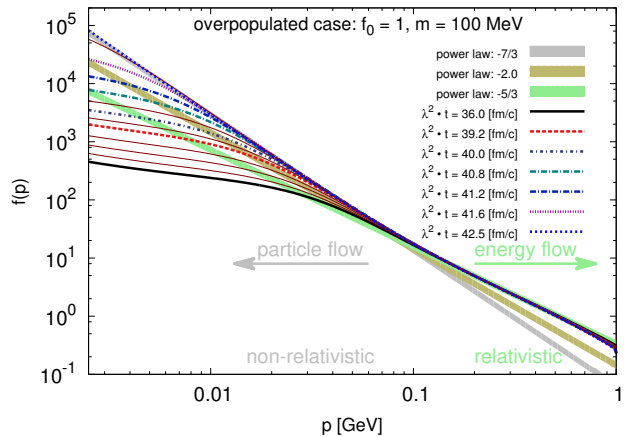}
\centering
\caption{Self-similar evolution with different power-law scalings of the
  one-particle distribution function in the region of relativistic and
  non-relativistic momenta, thin lines indicate intermediate
  propagation.}
 \label{fig:309a}
\end{figure}
Fig. \ref{fig:309a} summarizes the evolution (resolved in time) for the
overpopulated system with the initial state $f_0=1$, starting from
$\lambda^2t= 36\;\text{fm}/c$ up to the time at which the one-particle
distribution function reaches its maximum. Obviously, one observes two
regions with almost linear scalings in a double logarithmic plot. We
denote the momentum range with $p\ll m$ as non-relativistic, whereas
$p\gg m$ marks the relativistic regime.  In fact, both exponents can be
understood from examination of the scaling behavior of the collision
integral.  Therefore, we consider the individual terms of the
Boltzmann equation \eqref{eq:200} in App.\ \ref{sec:b} under
momentum rescaling.

To obtain a feeling for particle and energy transport, we also plotted the differential particle 
as well as energy density in Figs.\ \ref{fig:309b} and \ref{fig:309c} for an isotropic system with respect 
to the momentum $p$, by defining them as
\begin{equation}
\begin{split}
  \left(\frac{\dd n}{\dd p}\right)_{\text{part}}
  &:=\frac{p^2f_{\text{part}}\left(t,p\right)}{2\pi^2n_{\text{tot}}}\,,
  \\
  \left(\frac{\dd \epsilon}{\dd p}\right)_{\text{part}}
  &:=\frac{Ep^2f_{\text{part}}\left(t,p\right)}{2\pi^2\epsilon_{\text{tot}}}\,.
\label{eq:306}
\end{split}
\end{equation}
In our approach with a finite particle mass the infrared region is of
non-relativistic character, so according to \eqref{eq:b12} the particle
cascade should scale like a power law $1/p^{\alpha_4}$ with the exponent
$\alpha_4=7/3$. Indeed, this is exactly the case as can be deduced from
Figs.\ \ref{fig:309a} and \ref{fig:309b}, supporting the picture of a
stationary turbulence scaling in the infrared as reported in
\cite{PhysRevLett.74.3093,Semikoz:1995rd}. Thereby our scaling exponent
is equivalent to $\alpha=7/6$, since the distribution function with
respect to the momentum $p$ scales like $1/p^{\alpha_4}\sim 1/E^\alpha$
for the classical dispersion relation.

In addition to the discussed non-relativistic and inverse particle
cascade, we also observe an energy flow to the ultraviolet region with a
different scaling behavior. For momenta $p \gg m=100\;\MeV$ the
relativistic dispersion relation scales almost linearly with increasing
momentum, and according to \eqref{eq:b13} one should expect
$\alpha_4=5/3$ to be the particular exponent of a direct energy
cascade. Obviously, at intermediate times there is an excess of the 
energy density over the equilibrium value in the momentum range 
$\left [0.1,\,1\right ]$ as seen in Fig.\ \ref{fig:309c}, which
explains the origin of the energy cascade towards the ultraviolet. The
stored energy in the infrared is negligibly small and is caused by a
finite value of the particle mass. Effectively, there is no energy
transport to the infrared and almost no particle transport to the
ultraviolet as expected\footnote{In contrast a significant energy
  transport to the ultraviolet and turbulent particle transport to the
  infrared can be observed}.  For $\lambda^2t= 44\;\text{fm}/c$ the
particle transport has already collapsed and the scaling is very close
to the equilibrium one, since $\alpha_4=7/3$ is not a root of the full
collision integral as given in \eqref{eq:203a}, including the
interaction between particles and the condensate. The direct energy
cascade still exists as seen from Fig.\ \ref{fig:309b}, even though the
condition \eqref{eq:b06} is only partially fulfilled for $f_0=1$ as the
initial distribution function.  Accordingly, for larger momentum scales
$p>1\;\GeV$ with $f\left(p\right)<1$ quantum effects become important
and dominate over the turbulent scaling behavior, leading to a
significantly different evolution of the one-particle distribution
function.
\begin{figure}
 \includegraphics[width=\columnwidth]{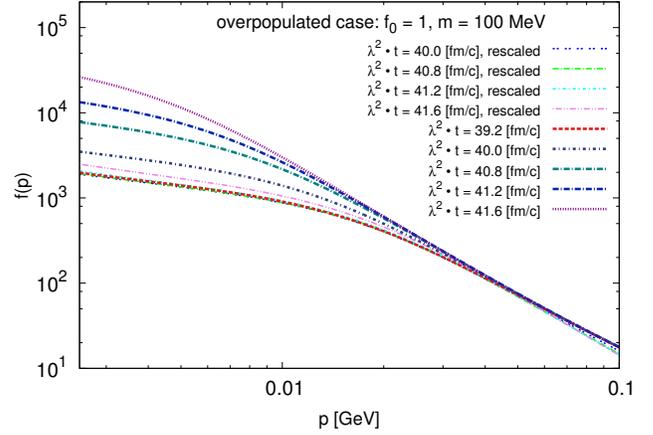}
\centering
\caption{Rescaled self-similar evolution of the one-particle
  distribution function in the in infrared region of the classical
  dynamics with $f\left(p\right)\gg 1$.}
 \label{fig:309d}
\end{figure}

Based on our numerical results the system undergoes a self-similar
evolution in the infrared for the inverse particle cascade and in the
ultraviolet for the direct energy cascade, thereby both exponents can be
explained by means of a stationary turbulent scaling. Nevertheless, at
this point we also refer to \cite{Berges:2012us, Orioli:2015dxa}, where the authors
study the universal self-similar dynamics by using classical-statistical
lattice simulations and vertex-resummed kinetic theory, emphasizing that
both are more reliable in the infrared sector of a strong turbulent
scaling, arising for $f\left(p\right)\geq 1/\lambda\gg 1$ in contrast to
a weak wave-turbulence, which exists in the region of
$1/\lambda\gg f\left(p\right)\gg 1$.

Since in our case the infrared region scales analogously to the
non-relativistic case and a time point $\tau_0$ exists at which the condition
\eqref{eq:b06} is surely fulfilled, we can apply the same ansatz for the
parametrization of the momentum distribution function as in
\cite{Semikoz:1995rd}. Following their approach, we separate $f(\tau,p)$
in an amplitude function $A\left(\tau\right)$ with an exponent $-\beta$
and the one-particle distribution function at $\tau_0$ with a rescaled
momentum $q:=p/A\left(\tau\right)$:
\begin{equation}
f\left(\tau,p\right)=A^{-\beta}\left(\tau\right)f\left(\tau_0, q\right)\,,
\label{eq:308}
\end{equation}
where $\tau$ is given by $\tau=t-\tau_0$. The amplitude function
$A\left(\tau\right)$ describes the scaling behavior of the occupation
number with respect to time. Inserting this ansatz in the isotropic
Boltzmann equation \eqref{eq:a11} without condensate, we derive in our
case the following form
\begin{equation}
A\left(\tau\right)=\left(\frac{\tau_c-\tau}{\tau_c}\right)^\frac{1}{2\left(\alpha_4-2\right)}
\label{eq:309}
\end{equation}
by requiring $A\left(0\right)=1$. Here,
$\lambda^2\left(\tau_c+\tau_0\right)\approx 43.2\;\text{fm}/c$ defines the time point at
which the distribution function for the zero mode grows to infinity
$f\left(t,p=0\right) \rightarrow \infty$. This value can be extracted
numerically by setting $n_\text{c}=0$ in \eqref{eq:a12} and so forbidding
the condensation process for all times. Furthermore, it turns out that
$\beta\equiv\alpha_4$. In Fig.\;\ref{fig:309d} we plotted the
one-particle distribution function at five different times, starting
with $\lambda^2t=39.2\;\text{fm}/c$, which determines $\tau_0$. In fact,
the rescaled distribution functions lie almost on top of each other as
expected from a self-similar evolution; only for the largest time
$\lambda^2t=41.6\;\text{fm}/c$ we observe a deviation due to the
condensation process which is not forbidden and starts already earlier,
modifying\footnote{The collision integral for the interaction with the 
condensate begins to grow rapidly.} the turbulent scaling as the
system approaches $\tau_c$.

\section{Summary}
\label{sec:4}

We have presented a numerical approach for treating the kinetics of
Bose-Einstein condensation by decomposing the one-particle phase space
distribution function into two parts. Thereby the zero-momentum part is
given by a product of the condensate density and a $\delta$-function in
momentum space. The non-zero momentum part contains the distribution of
excited modes. Inserting such an ansatz in the full Boltzmann equation
and discretizing the external momentum leads to a coupled set of
integro-differential equations for discrete modes. We have solved this
set of equations by using a quadrature scheme for the numerical
evaluation of the collision integrals and an adaptive Runge-Kutta scheme
for the time propagation of the differential equations. We compared our
results in the critical case to the partonic transport code BAMPS, which
applies a test-particle Monte-Carlo ansatz to solve the Boltzmann
equation. The time evolution of both simulations is in good agreement
with a much better performance for our code in the kinetic regime of a
relativistic Bose-Einstein condensate.

In the last Section we have initialized a strongly overpopulated system
of massive scalar bosons and have shown that our kinetic approach leads
to an accurate description of the equilibration process, evolving the
system exactly to the expected values for the temperature, chemical
potential and the condensate value. In this overpopulated case the time
evolution shows a typical particle cascade to the infrared momentum
region, developing in a self-similar way with a non-relativistic scaling
exponent in the inertial region of turbulent dynamics.  Here, the
occupation numbers of soft modes overshoot the equilibrium values by
several orders of magnitude, followed by a relaxation to the equilibrium
curve because of the condensation process.  In contrast the energy flow
to the ultraviolet region evolves with a relativistic exponent,
confirming a weak wave turbulence scaling.

In a future work we will include other types of matrix elements which
can be directly applied to the formation of a condensate like state of
gluons in the initial state of a heavy ion collision. Particularly, the
impact of the particle changing process $2\leftrightarrow 3$ could avoid
the formation of a condensate and has to be studied in detail. Therefore
we will also include a Bjorken expansion along the rapidity axis with an
anisotropic distribution function in $(2+1)$ dimensions. Furthermore,
the presented numerical approach is part of a code based on an on-shell
approximation of self-consistent quantum kinetic equations from the 2PI
effective action of the linear $\sigma$ model with constituent quarks
and will be used for the study of dynamical fluctuations during the
chiral phase transition.

\begin{acknowledgments}
A.\ M.\ acknowledges support from the
Helmholtz Research School for Quark Matter Studies (H-QM) and HIC for
FAIR. HvH.\ has been supported by the
Deutsche Forschungsgemeinschaft (DFG) under grant number GR 1536/8-1.
K.\ Z.\ was supported by the Helmholtz International
Center for FAIR within the framework of the LOEWE
program launched by the State of Hesse.
Numerical computations have been performed at the Center for
Scientific Computing (CSC).
\end{acknowledgments}

\appendix
\section{Isotropic Boltzmann equation}
\label{sec:a}

In the case of a $\phi^4$-interaction with
$\frac{1}{\nu}|\mathcal M_{12\rightarrow 34}|^2=18\lambda^2$ one can
integrate out the angular dependence in \eqref{eq:200} by simply using
the Fourier transform of the momentum conserving $\delta$-function
\begin{equation}
\begin{split} \delta^{(3)}\left (\vec{p}\right )&=\int_{\R^3}
\frac{\dd^3q}{\left (2\pi\right )^3}\exp\left (-\ii\vec{p}
\cdot\vec{q}\right )\\ &=\int_0^{\infty} \,\dd q \, q^2 \int_0^{2 \pi}
\, \dd \varphi \int_0^{\pi} \dd \vartheta_q\sin\vartheta_q \\ & \qquad
\exp\left [-\ii pq\cos\vartheta_q\right ]\\ &=\int_0^{\infty} \dd q \,
q^2\left [\frac{4\pi}{pq}\sin\left (pq\right )\right ]\,,
\label{eq:a02}
\end{split}
\end{equation}
leading to the isotropic form of the Boltzmann equation,
\begin{equation}
\begin{split}
\partial_t f_1=\frac{9\lambda^2}{8\pi^4}&\int_0^{\infty}\,\dd p_2\,\dd p_3\,\dd
p_4\,\dd q \frac{1}{E_{1}E_{2}E_{3}E_{4}} \\ & \times \delta\left
(E_1+E_2-E_3-E_4\right ) \frac{p_2p_3p_4}{p_1q^2}\\ &\times\sin\left
(p_1q\right )\sin\left (p_2q\right )\sin\left (p_3q\right )\sin\left
(p_4q\right )\\ &\times [\left (1+f_1\right )\left (1+f_2\right )f_3f_4
\\ & \qquad -f_1f_2\left (1+f_3\right )\left (1+f_4\right ) ]\,.
\label{eq:a03}
\end{split}
\end{equation}
Defining an auxiliary function which depends on the involved momenta
\begin{equation} \mathcal{F}_{1,2\leftrightarrow 3,4}^{p_1}=\int_0^{\infty}\,\dd q
\frac{\sin\left (p_1q\right )\sin\left (p_2q\right )\sin\left
(p_3q\right )\sin\left (p_4q\right )}{p_1q^2} \, ,
\label{eq:a04}
\end{equation}
and integrating out the $q$-dependence leads to
\begin{align}
\begin{split}
\mathcal{F}_{1,2\leftrightarrow 3,4}^{p_1=0}=\frac{\pi}{8}\Bigg[&\sign\left(p_2+p_3-p_4\right) \\
&-\sign\left(p_2-p_3-p_4\right) \\
&+\sign\left(p_2-p_3+p_4\right) \\
&-\sign\left(p_2+p_3+p_4\right)\Bigg]\,,
\label{eq:a05}
\end{split}\\
\begin{split}
\mathcal{F}_{1,2\leftrightarrow
3,4}^{p_1>0}=\frac{\pi}{16p_1} \Bigg[ &\left |p_1-p_2-p_3-p_4\right | \\
& -\left |p_1+p_2-p_3-p_4\right | \\ &-\left |p_1-p_2+p_3-p_4\right | \\
&+\left |p_1+p_2+p_3-p_4\right | \\ &-\left |p_1-p_2-p_3+p_4\right | \\
&+\left |p_1+p_2-p_3+p_4\right | \\ &+\left |p_1-p_2+p_3+p_4\right | \\
&-\left |p_1+p_2+p_3+p_4\right | \Bigg]\,,
\label{eq:a06}
\end{split}
\end{align}
where the form of $\mathcal{F}_{1,2\leftrightarrow
3,4}^{p_1}$ depends especially on the incoming momentum $p_1$.  The
remaining energy conserving $\delta$-function is used to reduce the
dimension of the integral by writing
\begin{equation} \delta\left (E_1+E_2-E_3-E_4\right
)=\frac{E_4}{p_4}\delta\left (p_4-\tilde p_4\right )\,.
\label{eq:a07}
\end{equation} Here $\tilde p_4$ is given via energy-momentum
conservation with the relativistic dispersion relation
\begin{equation} \tilde p_4=\sqrt{\left (E_1+E_2-E_3\right )^2-m_4^2}\,.
\label{eq:a08}
\end{equation} Finally, one ends up with the following expression for
the isotropic Boltzmann equation
\begin{equation}
\begin{split}
\partial_t f_1=\frac{9\lambda^2}{8\pi^4} & \int_0^{\infty}\,\dd p_2\int_0^{\infty}\,\dd
p_3\frac{p_2p_3}{E_1E_2E_3}\theta\left (\tilde p_4^2\right
)\mathcal{F}_{1,2\leftrightarrow 3,4}^{p_1} \\ &\times [\left
(1+f_1\right )\left (1+f_2\right )f_3f_4 \\ & \qquad -f_1f_2\left
(1+f_3\right )\left (1+f_4\right ) ]\,,
\label{eq:a09}
\end{split}
\end{equation} where the $\theta$-function ensures that $\tilde{p}_4^2
\geq 0$.

In analogy to \eqref{eq:a04} one can define the auxiliary functions
\begin{equation}
\begin{split} \mathcal{F}_{1,2\leftrightarrow 3,4}^{p_2=0}&=\int_0^{\infty}\,\dd q
\frac{\sin\left (p_1q\right )\sin\left (p_3q\right )\sin\left
(p_4q\right )}{p_1q}, \\ \mathcal{F}_{1,2\leftrightarrow
3,4}^{p_3=0}&=\int_0^{\infty}\,\dd q \frac{\sin\left (p_1q\right )\sin\left
(p_2q\right )\sin\left (p_4q\right )}{p_1q}, \\
\mathcal{F}_{1,2\leftrightarrow 3,4}^{p_4=0}&=\int_0^{\infty}\,\dd q
\frac{\sin\left (p_1q\right )\sin\left (p_2q\right )\sin\left
(p_3q\right )}{p_1q}\,.
\label{eq:a10}
\end{split}
\end{equation}
By means of \eqref{eq:a09} and \eqref{eq:a10} the evolution equation for
the distribution function \eqref{eq:203a} as well as the condensate
density \eqref{eq:204} can be written in their final isotropic form

\begin{align}
\begin{split}
\partial_t f(t,p_1)=\frac{9\lambda^2}{8\pi^4}&\int_0^{\infty}\,\dd p_2\int_0^{\infty}\,\dd
p_3\frac{p_2p_3}{E_1E_2E_3} \\ &\times \theta\left (\tilde p_4^2\right
)\mathcal{F}_{1,2\leftrightarrow 3,4}^{p_1>0}\\ &\times [\left
(1+f_1\right )\left (1+f_2\right )f_3f_4 \\ & \qquad -f_1f_2\left
(1+f_3\right )\left (1+f_4\right ) ]\\
+\frac{9\lambda^2}{4\pi^2}&\int_0^{\infty}\,\dd p_3\frac{p_3}{E_1E_2E_3}\theta\left
(\tilde p_4^2\right )\mathcal{F}_{1,2\leftrightarrow 3,4}^{p_2=0}\\
&\times n_{\text{c}}\left [f_3f_4-f_1\left (1+f_3+f_4\right )\right ]\\
+\frac{9\lambda^2}{4\pi^2}&\int_0^{\infty}\,\dd p_2\frac{p_2}{E_1E_2E_3}\theta\left
(\tilde p_4^2\right )\mathcal{F}_{1,2\leftrightarrow 3,4}^{p_3=0}\\
&\times n_{\text{c}}\left [\left (1+f_1+f_2\right )f_4-f_1f_2\right ]\\
+\frac{9\lambda^2}{4\pi^2}&\int_0^{\infty}\,\dd p_2\frac{p_2}{E_1E_2E_4}\theta\left
(\tilde p_3^2\right )\mathcal{F}_{1,2\leftrightarrow 3,4}^{p_4=0}\\
&\times n_{\text{c}}\left [\left (1+f_1+f_2\right )f_3-f_1f_2\right ]\,,
\label{eq:a11}
\end{split}\\
\begin{split}
\partial_t n_{\text{c}}\left (t\right
)=\frac{9\lambda^2}{8\pi^4}&\int_0^{\infty}\,\dd p_2\int_0^{\infty}\,\dd
p_3\frac{p_2p_3}{E_1E_2E_3} \\ &\times \theta\left (\tilde p_4^2\right
)\mathcal{F}_{1,2\leftrightarrow 3,4}^{p_1=0}\\ &\times
n_{\text{c}}\left [f_3f_4-f_2\left (1+f_3+f_4\right ) \right ]\,.
\label{eq:a12}
\end{split}
\end{align}

\section{Turbulent scaling}
\label{sec:b}

The derivations in this section are based on methods and ideas described
in \cite{Micha:2004bv}.  The relativistic integration measure of the
Boltzmann equation \eqref{eq:200} has the form
\begin{equation}
\begin{split}
\dd\Omega:=&\frac{\left(2\pi\right)\delta^{(4)}\left(P_1+P_2-P_3-P_4\right)}{2\omega\left(\vec p_1\right)}\\
&\times\frac{\left |\mathcal{M}_{12\rightarrow 34} \right|^2}{\nu}\prod_{i=2}^{4}\frac{\dd^3\vec p_i}{2\omega\left(\vec p_i\right)}\,,
\label{eq:b01}
\end{split}
\end{equation}
where $\omega\left(\vec p\right)$ is given by the energy-momentum dispersion relation.
For an isotropic and homogeneous system this relation 
depends only on the modulus $p$ of the momentum vector $\vec p$ and scales like
\begin{equation}
\begin{split}
\omega\left(p\right)&=s^{-z}\omega\left(sp\right)\\
&\simeq 
\begin{cases}
s^{-1}\omega\left(sp\right)\quad\text{for $p\gg m$ (relativistic)} \\
s^{-2}\omega\left(sp\right)\quad\text{for $p\ll m$ (non-relativistic)}\end{cases}
\label{eq:b02}
\end{split}
\end{equation}
with the dynamic exponent $z$, describing the scaling behavior of the energy $\omega(p)$ with respect to $p$.
With the definition \eqref{eq:b01} the collision integral of non-zero modes can be written as 
\begin{equation}
\begin{split}
\mathcal C_{2\text{p}\leftrightarrow 2\text{p}}\sim\int\dd\Omega\,
[&\left (1+f_1\right )\left (1+f_2\right )f_3f_4 \\
&-f_1f_2\left (1+f_3\right )\left (1+f_4\right ) ]\,.
\label{eq:b03}
\end{split}
\end{equation}

Focusing only on systems with a homogeneous scaling of the integration measure, one obtains 
\begin{equation}
\dd\Omega\left(sp_1,...\,,sp_4\right)=s^{\nu_4}\dd\Omega\left(p_1,...\,,p_4\right)\,,
\label{eq:b04}
\end{equation}
where $\nu_4$ denotes the scaling exponent of $\Omega$. The index refers
to the structure of the vertex with four interacting particles. From
considering the integration measure in the relativistic \eqref{eq:b01} and non-relativistic limit,
where terms proportional to $1/\omega\left(p\right)$ become almost constant
$\sim 1/m$, so that effectively only the energy-$\delta$ function contributes
with the exponent $-2$ to the overall scaling behavior, $\nu_4$ can be derived to be
\begin{equation}
\nu_4=
\begin{cases}
3d-\left(3+1\right)-4z=1\quad\text{(relativistic)}, \\
3d-\left(3+2\right)=4\quad\text{(non-relativistic)}\,,
\end{cases}
\label{eq:b05}
\end{equation}
with $d\equiv 3$ denoting the number of dimensions. The term in
brackets refers to an effective reduction of the dimensionality for the
integration measure due to momentum and energy conservation. 

Taking into account the large occupation number for the one-particle
distribution function $f\gg 1$ in the inertial region of turbulent
dynamics the Bose enhancement can be neglected and it follows
\begin{equation}
\begin{split}
F&:=[\left (1+f_1\right )\left (1+f_2\right )f_3f_4-f_1f_2\left (1+f_3\right )\left (1+f_4\right ) ]\\
&\simeq [\left(f_1 + f_2\right) f_3f_4-f_1f_2\left (f_3+f_4\right ) ]\,,
\label{eq:b06}
\end{split}
\end{equation}
so in this regime only terms of the form $f^3$ contribute significantly
to the integral, reducing the scaling complexity of the
integral. Following the assumption of a power-law $f\sim 1/p^{\alpha_4}$
in the inertial region of the evolution, one obtains a homogeneous
relation for the expression $F$ with respect to the occupation number,
respectively the momenta $p_1,...\,,p_4$.
\begin{equation}
F\left(sp_1,...\,sp_4\right)=s^{-3\alpha_4}F\left(p_1,...\,p_4\right)\,.
\label{eq:b07}
\end{equation}
This leads to a scaling exponent $\mu_4:=\nu_4-3\alpha_4$ for the full collision integral
\eqref{eq:b03},
\begin{equation}
\mathcal C_{2\text{p}\leftrightarrow 2\text{p}}\left[F\right]\left(sp_1\right)
= s^{\mu_4}\mathcal C_{2\text{p}\leftrightarrow 2\text{p}}\left[F\right]\left(p_1\right)\,,
\label{eq:b08}
\end{equation}
where it is pointed out that $C_{2\text{p}\leftrightarrow 2\text{p}}$ is a functional with respect to $F$ and
depends explicitly on the external momentum $p_1$.

In the next step, one includes conservation laws, more precisely
particle and energy densities by using the relations given in
\eqref{eq:303}, but integrating only up to a finite value of the
external momentum $p_1$.  In an isotropic and homogeneous system the
time evolution of the particle density in a momentum interval
$\left[0,p\right]$ reads
\begin{equation}
\begin{split}
\partial_t n_{\text{part}}\left (t,p\right )&:=\int_{0}^p\frac{\dd p_1
p_1^2}{2\pi^2}\,\partial_t f_{\text{part}}\left(t,p_1\right)\\
&=\int_{0}^p\frac{\dd p_1 p_1^2}{2\pi^2}
\mathcal C_{2\text{p}\leftrightarrow 2\text{p}}[F]\left(p_1\right)\,.
\label{eq:b09}
\end{split}
\end{equation}
Consequently, this expression is proportional to the isotropic particle
flux in or out of a sphere with radius $p$ and has to be constant as
well as independent of the momentum-scale in the region of stationary
turbulence.  That means a rescaling of the argument in \eqref{eq:b09}
leads to
\begin{equation}
\partial_t n_{\text{part}}\left (t,p\right )=
\int_{0}^p\frac{\dd p_1 p_1^2}{2\pi^2}s^{-\mu_4}
\mathcal C_{2\text{p}\leftrightarrow 2\text{p}}[F]\left(sp_1\right)\,.
\label{eq:b10a}
\end{equation}
In analogy to the particle density one obtains the following relation
for the energy density:
\begin{equation}
\partial_t\epsilon_{\text{part}}\left(t,p\right)=
\int_0^p\frac{\dd p_1 p_1^2}{\left(2\pi\right)^3}s^{-z-\mu_4}\omega\left(sp_1\right)\,
\mathcal C_{2\text{p}\leftrightarrow 2\text{p}}[F]\left(sp_1\right)\,.
\label{eq:b11a}
\end{equation}
Since the physics has to be scale invariant, one is free to choose $s=1/p_1$, resulting in
\begin{equation}
\partial_t n_{\text{part}}\left (t,p\right )\sim\frac{p^{d+\mu_4}}{d+\mu_4}C_{2\text{p}\leftrightarrow 2\text{p}}[F]\left(1\right)
\label{eq:b10b}
\end{equation}
for the particle as well as
\begin{equation}
\partial_t \epsilon_{\text{part}}\left (t,p\right )\sim\frac{p^{d+z+\mu_4}}{d+z+\mu_4}\omega\left(1\right)
C_{2\text{p}\leftrightarrow 2\text{p}}[F]\left(1\right)\,.
\label{eq:b11b}
\end{equation}
for the energy flux.
 
Due to the conservation of the particle density a scale invariant solution requires 
according to relation \eqref{eq:b10b} the following scaling behavior of the one-particle distribution function,
\begin{equation}
\begin{split}
& d+\mu_4 = d+\nu_4-3\alpha_4\stackrel{!}{=}0 \\
& \Rightarrow\alpha_4=
\begin{cases}
4/3\quad\text{(relativistic)}\\
7/3\quad\text{(non-relativistic)}\,.
\end{cases}
\label{eq:b12}
\end{split}
\end{equation}
Similarly, the exponent due to the conservation of the energy density reads
\begin{equation}
\begin{split}
& d+z+\mu_4 = d+z+\nu_4-3\alpha_4\stackrel{!}{=}0 \\
& \Rightarrow\alpha_4=
\begin{cases}
5/3\quad\text{(relativistic)}\\
8/3\quad\text{(non-relativistic)}\,.
\end{cases}
\label{eq:b13}
\end{split}
\end{equation}
From Eqs.\ \eqref{eq:b10b} and \eqref{eq:b11b} one immediately
recognizes that the scaling conditions \eqref{eq:b12}, \eqref{eq:b13}
for a stationary turbulence imply roots of a first degree for the
collision integral, otherwise the fluxes could not have a constant and
finite value.

\bibliography{becbib}
\end{document}